# On the problem of comparing Covid-19 fatality rates

Fabio Miletto Granozio, CNR-SPIN, Napoli, IT, email: fabio.miletto@spin.cnr.it

**Abstract**

Understanding Covid-19 lethality and its variation from country to country is essential for supporting governments in the choice of appropriate strategies. Adopting correct indicators to monitor the lethality of the infection in the course of the outbreak is a crucial issue. This works highlights how far the case fatality ratio, as calculated day by day in the course of the outbreak, is a misleading indicator for estimating the fatality rate, even if our attention is only restricted to the subset of confirmed cases. We prove that the final case fatality ratio for several major European countries is bound to largely exceed 10%. The difference in the case fatality ratio values between Italy and other major European countries considered in this work (except for the case of Germany), often interpreted in terms of peculiarities of the Italian population or health system, is to be mostly attributed to the more advanced stage of the Italian epidemic.

**Introduction**

Different countries are adopting different strategies to face the Covid-10 pandemic[1]. Monitoring Covid-19 lethality and its variation from country to country is essential for supporting governments in selecting the appropriate decisions. The adoption of correct indicators to monitor the lethality of the infection in the course of the outbreak is therefore a crucial issue. [2,3]

The following terms are used in this work: the *infection fatality ratio, IFR,* indicates the ratio between the total deaths and total infected patients within a given population; the *final case fatality ratio, $CFR^{fin}$*, indicates the ratio of confirmed deaths attributed to the infection and confirmed cases, as calculated after all patients deceased or recovered; *time dependent case fatality ratio* $CFR(t)$ to indicate the day by day $\frac{d_c(t)}{cc_c(t)}$ ratio calculated. Here, *t* is a discrete variable expressed in days, $d_c(t)$ the cumulative number of deaths at date *t*, $cc_c(t)$ the cumulative number of confirmed cases in date *t*. Calculating *IFR* is a formidable task, because of two errors that can hardly be avoided: *(a)* estimating *IFR* exactly from $CFR^{fin}$ would require registering, or correcting estimating, all Covid-19 infections and deaths in a country; *(b)* estimating $CFR^{fin}$ itself in the course of an outbreak, as better clarified in the following, is a tricky and non-trivial operation. Large and efficient testing programs allow accounting for non-symptomatic patients, thus increasing $cc_c(t)$. This lowers the error *(a)*, i.e. lowers $CFR^{fin}$ to values closer to *IFR*. This work is instead dedicated to the evaluation of error *(b)*.

**Data analysis**

Tab. 1 reports on the $CFR(t)$ values updated on April 5, for eight countries, chosen among those currently hit harder by the pandemic: United States of America (US), Italy (IT), Spain (ES), Germany (DE), France (FR), United Kingdom (UK), The Netherlands (NL), Korea (KR).

| Country | IT | ES | NL | UK | FR | US | KR | DE |
|---|---|---|---|---|---|---|---|---|
| CFR | 12,30% | 9,60% | 9,90% | 10,30% | 8,70% | 2,90% | 1,80% | 1,60% |

Tab. 1: $CFR(t)$ values updated on April 5 for KR, IT, ES, US, DE, UK, NL, and FR.

In order to show how deceptive, though common, is the approach of ignoring the error (b), thus directly estimating $CFR^{fin}$ from $CFR(t)$ in a course of an outbreak, we initially resort to a simulation. The set of curves in Fig. 1a, cumulative confirmed cases ($cc_c$), cumulative deaths ($d_c$), total active cases ($act_t$), and cumulative recovered cases ($rc_c$) was derived by implementing an epidemic model based on very simple and general assumptions and explained in [4], where simulations can be run as a function of the input parameters. We focus on two of such parameters: $CFR^{fin}$, set to 10%, and the average diagnosis-to-death delay time $T_D = 8$ days. Based on the $cc_c(t)$ and $d_c(t)$ curves from Fig. 1b, we calculate the $CFR(t)$ curve of the simulated epidemic and plot it in Fig. 2 (dark blue line).

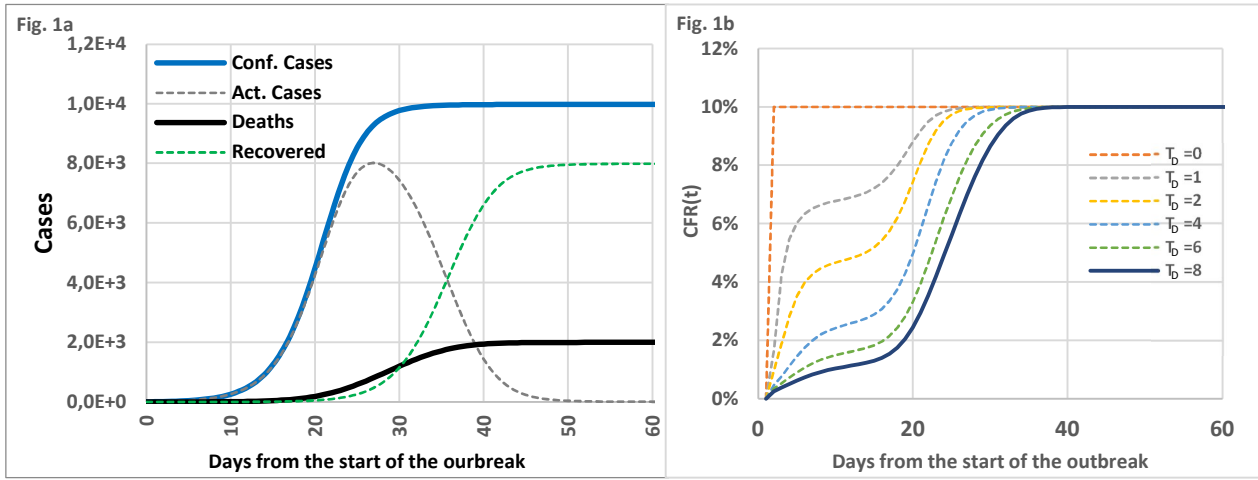

**Fig. 1:** (a) Cumulative confirmed cases ($cc_c$), cumulative deaths ($d_c$), total active cases ($ac_t$), and cumulative recovered cases ($rc_c$) plotted versus time for a standard simulation of an epidemic. The $CFR(t)$ value is set to 10%, and the average diagnosis-to-death delay time $T_D$ to 8 days; (b) Case fatality ratio $CFR(t)$ plotted vs. time for the simulated epidemic in (a) (full blue line) and compared with curves obtained for different delay times $T_D$ (dotted lines).

Dotted curves obtained setting shorter $T_D$ delay times are shown for comparison, confirming that the $T_D$ value determines the mismatch between $CFR(t)$ and $CFR^{fin}$. We deduce by the comparison that before the peak of active cases $CFR(t)$ is a underestimates $CFR^{fin}$ by a large factor. In particular, for $T_D = 8$ days, the $\frac{CFR(t)}{CFR}$ ratio remains below 20% for a substantial portion of the outbreak.

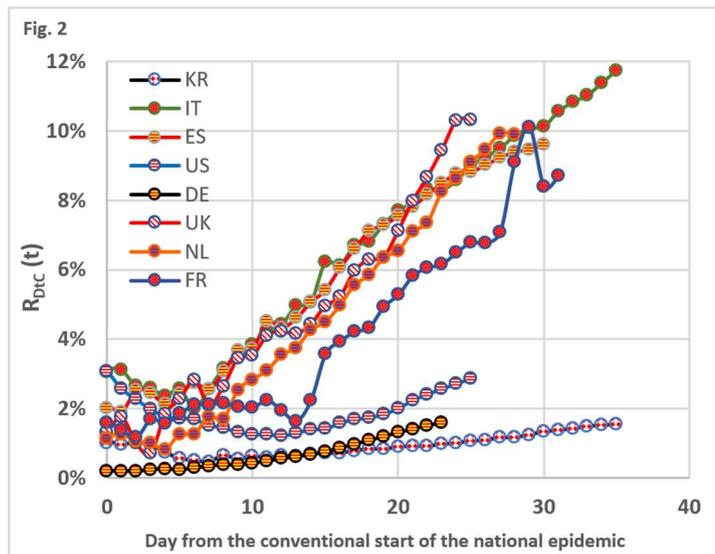

**Fig. 2:** Case fatality ratio $CFR(t)$ plotted vs. time for KR, IT, ES, US, DE, UK, NL, and FR. A relative shift of the time axes was applied for alignment, based on the evolution of deaths in each country.

The plot of the $CFR(t)$ curves for all the nations addressed in Tab. 1 is shown in Fig. 3. The curves have been shifted on the time axis for better alignment, based on the evolution of the death curve in the country. The earliest days are omitted because largely affected, for most nations, by insufficient testing. The plot confirms the claims based on the simulation: all the curves show a clear growing trend and a behaviour which is reminiscent of the dark blue curve in Fig. 2. Both IT and DE presently show a $CFR(t)$ value exceeding by over five times the ones of the early stage of the epidemic, while a slightly lower but comparable factor applies to UK, ES, NL and FR. The curves for IT, ES, UK and NL seem to be all bound to converge towards comparable final $R_{CF}$ values, in spite of the large present differences. The behaviour of the US curve is so far very difficult to interpret and certainly affected by large State-to-State inhomogeneities, also in terms of starting date of the outbreak.

**Conclusions**

Both the simulation and the examples above clarify that scientists and policymakers should beware of $CFR(t)$ as a misleading indicator for calculating the infection fatality rate in the course of the outbreak, *even if our attention is only restricted to the subset of confirmed cases*. More comprehensive data analysis methods based on the estimate of the average diagnosis-to-death delay time $T_D$ are needed. The largely discussed[5] difference in the $CFR(t)$ values between Italy and other major European countries addressed in this work, except for the case of Germany, are rather to be attributed to the more advanced stage of the Italian epidemic. The $CFR^{fin}$ for IT, ES, FR, UK, NL are expected to be grossly comparable at the end of the epidemic, all well above 10%. Germany is converging to lower $CFR^{fin}$. Probably larger and more efficient testing programs contribute to this result.